\documentclass[lettersize,journal]{IEEEtran}
\def\BibTeX{{\rm B\kern-.05em{\sc i\kern-.025em b}\kern-.08em
		T\kern-.1667em\lower.7ex\hbox{E}\kern-.125emX}}
\usepackage[caption=false,font=normalsize,labelfont=sf,textfont=sf]{subfig}
\usepackage{amsmath,graphicx}
\usepackage{amsmath,amsfonts,amssymb}
\usepackage{xcolor}
\usepackage{algorithmicx,algorithm}
\usepackage{algpseudocode}
\usepackage{bbding}
\usepackage{tabularx}
\usepackage{multirow}
\usepackage{booktabs}
\usepackage{makecell} 

\title{OL-DN: Online learning based dual-domain network for HEVC intra frame quality enhancement}

\author{Renwei Yang, Shuyuan Zhu, \IEEEmembership{Member,~IEEE}, Xiaozhen Zheng, and Bing Zeng, \IEEEmembership{Fellow,~IEEE}
	
	\thanks{R. Yang, S. Zhu and B. Zeng are with School of Information and Communication Engineering, University of Electronic Science and Technology of China, Chengdu, China.}
	\thanks{X. Zheng is with SZ DJI Technology Co., Ltd., Shenzhen, China}
	\vspace{-5mm}
}

\begin{document}
%
\maketitle
\begin{abstract}
Convolution neural network (CNN) based methods offer effective solutions for enhancing the quality of compressed image and video. However, these methods ignore using the raw data to enhance the quality. In this paper, we adopt the raw data in the quality enhancement for the HEVC intra-coded image by proposing an online learning-based method. When quality enhancement is demanded, we  online train our proposed model at encoder side and then use the parameters to update the model of decoder side. This method not only improves model performance, but also makes one model adoptable to multiple coding scenarios. Besides, quantization error in discrete cosine transform (DCT) coefficients is the root cause of various HEVC compression artifacts. Thus, we combine frequency domain priors to assist image reconstruction. We design a DCT based convolution layer, to produce DCT coefficients that are suitable for CNN learning.
Experimental results show that our proposed online learning based dual-domain network (OL-DN) has achieved superior performance, compared with the state-of-the-art methods.
\end{abstract}

\begin{IEEEkeywords}
	Chrominance, compression, quality enhancement, HEVC, CNN
\end{IEEEkeywords}

\section{Introduction}
Over the past few years, deep learning based approaches are became popular in video coding research \cite{lei2021deep,li2021high,liu2021fast}, among which convolution neural network-based  methods have demonstrated impressive performance on enhancing quality of HEVC coded frames. A variable-filter-size CNN with residual learning (VR-CNN) is proposed in \cite{dai2017vrcnn} to remove HEVC compression artifacts. Frame-enhancement CNN (FE-CNN) proposed in \cite{li2018fecnn} employs long and short skip connections to improve frame quality. Multi-frame guided attention network (MGANet) \cite{meng2021robust} enhances video quality based on long-short-term frame dependency and coding unit boundaries. Content-aware CNN (CA-CNN) \cite{jia2019contentaware} contains multiple models and selects the most appropriate one to enhance each coding tree unit. Adaptive-switching network (ASN) \cite{lin2020partitionaware} switches to different enhancement models according to frame content. A network adopting recursive design and residual learning (RR-CNN) is proposed in \cite{zhang2020rrcnn} as HEVC intra frame post-processing filter. Video coding prediction modes and units partition map are utilized as side information in frame-wise quality enhancement CNN (FQE-CNN) \cite{huang2021fqecnn} to boost quality enhancement of HEVC coded frames.

Different from other image restoration tasks, the raw data is available at the encoder side in image and video coding task. However, these existing CNN-based methods neglect employing it to enhance coded frame quality. In this work, we use the raw frame at encoder as ground-truth to online train OL-DN to overfit to this frame. Then, we transmit the new parameters to update the OL-DN at decoder. Therefore, the model gains high enhancement performance, as well as being adoptable to various coding scenarios. Moreover, we design a lightweight adaptive layers (AL) as the only online trainable part to reduce online training complexity. AL implements channel attention similarly to SE block \cite{hu2019senet}, but with less computation complexity. In OL-DN, we combine wide activation \cite{fan2018wdsr} and channel attention mechanism to design the wide block (WB) as an effective feature extraction block. And furthermore, we use AL to implement channel attention in WB to obtain online learning based wide block (OL-WB).

Besides, since quantization error in discrete cosine transform (DCT) coefficients is the root cause of various quality degradation in HEVC, Guo et al. \cite{guo2016ddcn} and Zhang et al. \cite{zhang2018dmcnn} introduce the priors of DCT coefficients as side information to remove the JPEG artifacts. Nevertheless, the strength of CNN lies in learning the inter dependencies between adjacent elements, while DCT coefficients have weak correlation with adjacent ones. Therefore the frequency coefficients are sub-optimal for CNN learning and frequency priors are not efficiently extracted in those methods. Therefore, we propose a DCT-based convolution layer (DCT-conv), which clusters DCT coefficients of the same frequency spectrum into one channel, and remains their relative positional relationship in spatial domain consistent with that in frequency domain. DCT-conv strengthens the frequency coefficients correlation and makes them more compatible with CNN learning. Consequently, our method can effectively extract frequency priors to improve quality enhancement performance.

The coding information was also adopted in the CNN-based quality enhancement solution. The coding unit mask was defined in \cite{huang2021fqecnn} as the prior information for the CNN-based post-processing. Also, the transform unit partition map was employed in  \cite{meng2021robust} to improve the quality of video frames. Besides the coding information, the channel correlation between the luma image (Y) and the chrom images (U and V) may also be utilized to improve the quality of compressed images.However, most of the existing CNN-based methods process each channel independently, which ignores the channel correlation and cannot achieve a high efficiency. In this work, we design the deep network in which we use the luma component as the guidance to restore the chroma component in both spatial and frequency domains.

\begin{figure*}[t]
	\vspace{-1em}
	\centering
	\includegraphics[width=0.7\textheight]{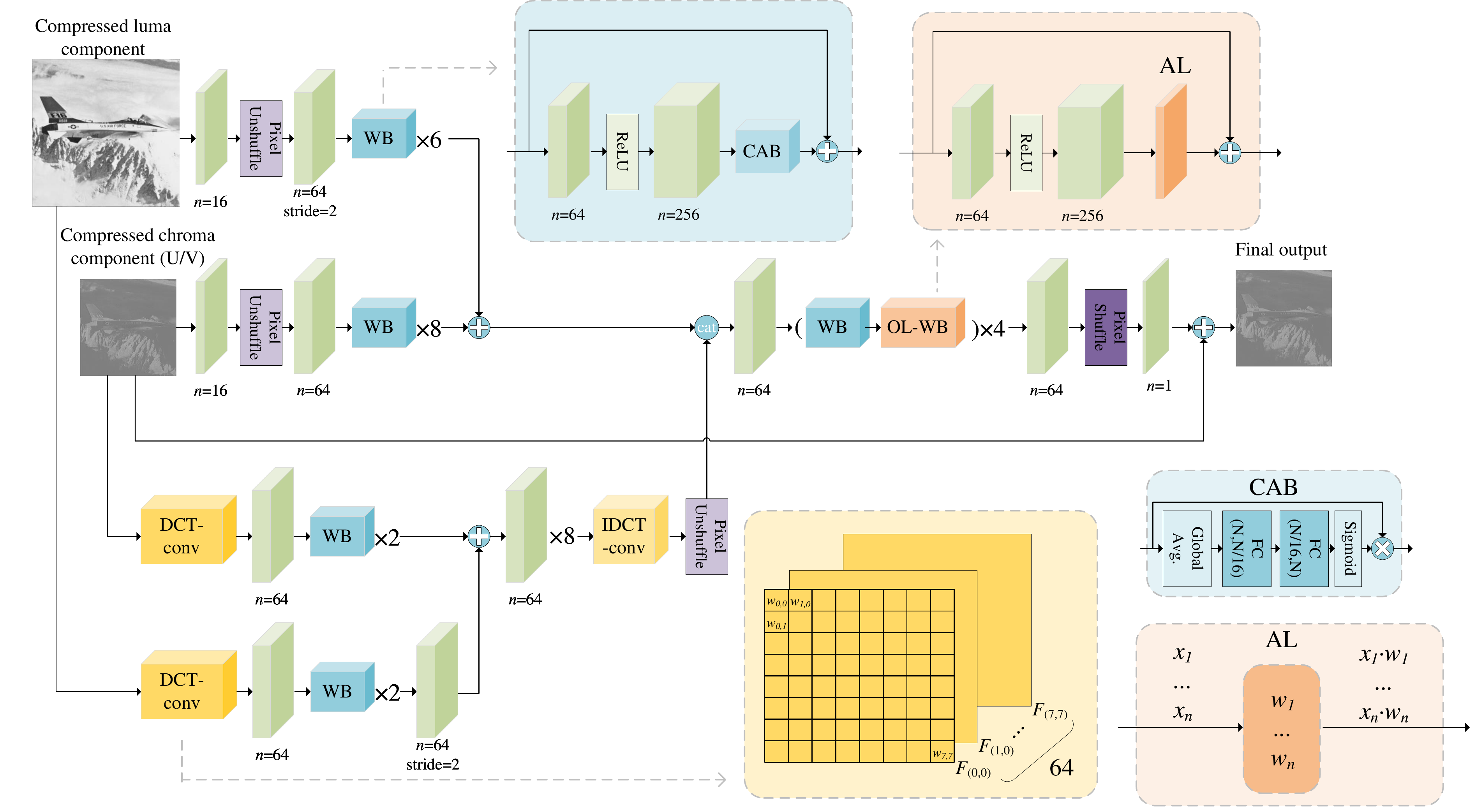}
	{\small{\caption{The framework of our proposed OL-DN. Green layer indicates 3$\times$3 convolution layer with $n$ output channels.}\label{f:network}}}
	\vspace{-.5em}
\end{figure*}

\begin{figure}[h]
	\vspace{-.em}
	\centering
	\includegraphics[width=0.35\textheight]{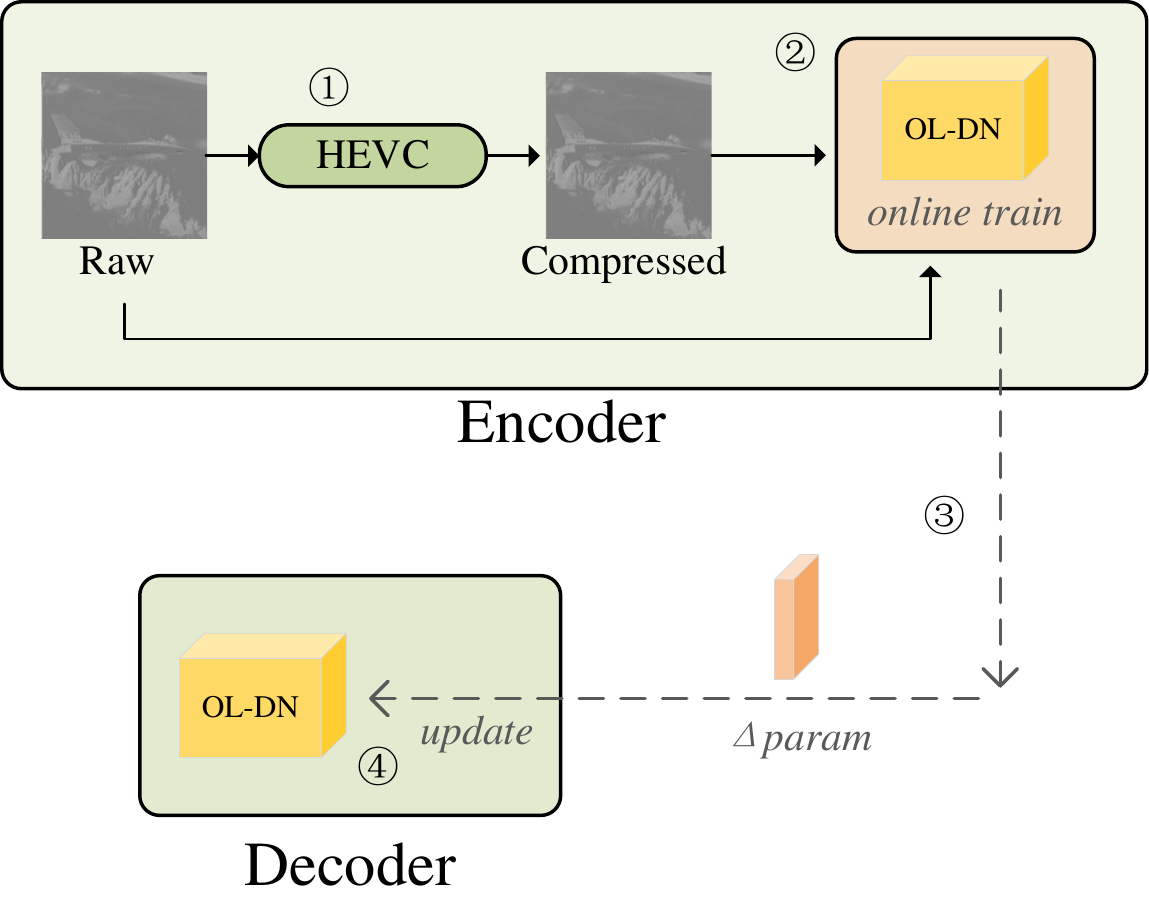}
	\vspace{-1.em}
	{\small{\caption{The flow chart of our online learning method.}\label{f:olearn}}}
\end{figure}

\section{Proposed Method}
%
%

\subsection{Online Learning}
\label{sec:ol}
Our proposed online learning method updates the OL-DN to gain performance improvement, and the its procedure is summarized in Fig. \ref{f:olearn}. Specifically, an OL-DN model is offline trained as the baseline model and it is arranged at both encoder and decoder sides. During the practical coding, OL-DN is online trained at the encoder side with raw data as the ground-truth image. Then, the online trained model parameters are coded via Huffman coding and transmitted to decoder side. Finally, the received parameters update the baseline model at decoder side. Note that to limit the extra encoder complexity and bit-rate for online learning, only the parameters of the designed adaptive layer (AL) are updated during online learning, and only the residual of parameters between the baseline model and online trained model is transmitted.

Online learning enables the model to learn the mapping from input to desired output directly, bringing considerable performance improvement to OL-DN. Therefore, OL-DN does not require either extremely deep network or multiple offline trained models to can outperform previous methods.

\subsection{Network Architecture of OL-DN}
\noindent
\textbf{Overview:} The architecture of OL-DN is illustrated in Fig. \ref{f:network}. It takes luma and chroma inputs, and outputs enhanced chroma image. OL-DN mainly consists of wide block (WB), online learning-based wide block (OL-WB), and DCT-based convolution layer (DCT-conv). We employ WB to extract input features and reconstruct output image. 
In reconstruction process, we insert adaptive layer (AL) in WB to obtain OL-WB, allowing the network online learnable. The DCT-conv converts image from pixel domain into frequency domain, and IDCT-conv conducts the inverse process. In OL-DN, the pixel shuffle and unshuffle  layers are utilized to reduce computation complexity\cite{shi2016pixelshuffle}. Moreover, to avoid the exploding and vanishing gradient problems we apply skip connections \cite{he2016residual} between inputs and outputs of OL-DN, WB, and OL-DB.

\noindent
\textbf{Wide block:} WB contains two wide-activated convolution layers and a channel attention block (CAB). We expand WB input channels before ReLU activation, to achieve better model performance for given computation complexity\cite{fan2018wdsr}.
Then, CAB implements channel attention mechanism to boost the learning ability of network \cite{hu2019senet}. A CAB input $\in \mathbb{R} {^{N \times H \times W}}$ is globally averaged to a 1-D vector and passes through two fully connected layers and sigmoid activation function, then it re-calibrates the channel weights of CAB input via multiplication, producing the CAB output. 

Let ${conv}^k_n$ denote the convolutional layer with $k \times k$ kernel and $n$ output channels. With input $x$, the output of WB is obtained by

\begin{equation}\label{Eq.wb}
y = x + CAB\left( {conv_{64}^3({\mathop{\rm Re}\nolimits} LU(conv_{256}^3(x)))} \right)
\end{equation}


\noindent
\textbf{Online learning-based wide block:}
We substitute AL for CAB in WB to obtain OL-WB. Our proposed AL consists of $N$ weighting parameters $\{w_1,\ldots,w_N\}$, which are corresponding to $N$ channels of the input $X$, where $X=\{x_1,\ldots,x_N\}$ and $x_i$ is the $i$-th channel of $X$. The output of AL is obtained by $AL(X) = \{ x_1 \cdot w_1,\ldots,x_N \cdot w_N \}$. Its lightweight structure guarantees low time complexity for online learning, and low bit-rate for transmitting the parameter difference. Since the neural network channel weights might be similar at near depths, we employ OL-WB at intervals to maximize the efficiency of online learning. 

Compared to CAB, AL also implements channel attention via channel-wise multiplication but its re-calibration parameters are directly online learned from the current input. Thus, it can achieve more accurate re-calibration for channels to improve the learning ability of the network. In addition, it can reduce computational complexity for the model. 

\begin{figure}[h]
	\vspace{-.em}
	\centering
	\includegraphics[width=0.35\textheight]{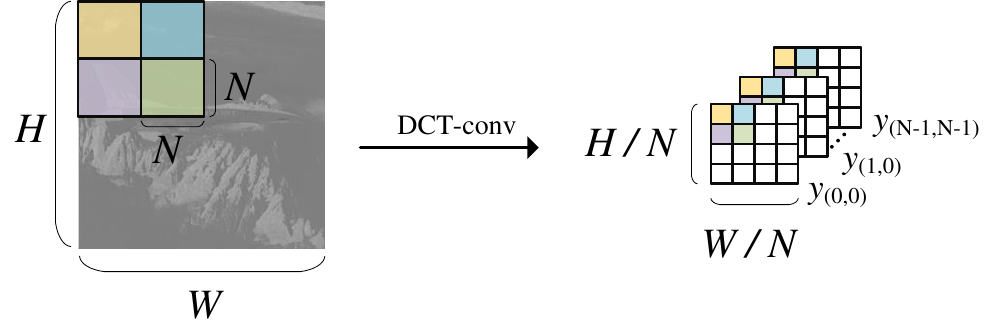}
	\vspace{-1.em}
	{\small{\caption{The process of DCT-conv converting image into frequency domain.}\label{f:dct}}}
	\vspace{-.5em}
\end{figure}

\noindent
\textbf{DCT-based convolution layer:} DCT-conv converts image from pixel domain to frequency domain and IDCT-conv accomplishes the inverse process. More specifically, DCT-conv consists $N^2$ kernels sliding over input image in a non-overlapped manner, with each kernel corresponding to a frequency spectrum. To obtain a DCT coefficient $y_{(u,v)}$, the corresponding DCT-conv kernel $F_{(u,v)}$ contains $N^2$ weights

\begin{equation}\label{Eq.dct}\scalebox{1.0}
	{$
	{w_{(i,j)}} = c(u)c(v)\cos \left( {\frac{{\left( {i + 0.5} \right)\pi }}{N}u} \right)\cos \left( {\frac{{\left( {j + 0.5} \right)\pi }}{N}v} \right)
	$}
\end{equation}

\noindent
where $u,v,i,j \in \{0,...,N-1\}$, and $c(u)={\sqrt {1/N}}$ if $u=0$ otherwise $c(u)={\sqrt {2/N}}$. Moreover, $N=8$ in this work. The process of DCT-conv is illustrated in Fig. \ref{f:dct}. For a given image $x\in \mathbb{R} {^{1 \times H \times W}}$, the DCT-conv outputs $y\in \mathbb{R} {^{N^2 \times (H/N) \times (W/N)}}$. Each $N\times N$ block in $x$ is converted into $N^2$ DCT coefficients in output channels from $y_{(0,0)}$ to $y_{(N-1,N-1)}$, representing frequencies from low to high. The corresponding pixel blocks and frequency coefficients are marked with the same color in Fig. \ref{f:dct}.
In conventional block-wise DCT, adjacent coefficients represent amplitudes of different frequency spectrum, so they are weakly correlated especially at block boundaries. In contrast, our proposed DCT-conv clusters coefficients of the same frequency spectrum into one channel, and adjacent coefficients also correspond to block neighbors in spatial domain. Therefore, we have strengthen elements correlation to match CNN's ability in learning local relationship, allowing the model to effectively extract frequency priors to improve quality enhancement performance.

\noindent
\textbf{Y guidance:}In color video compression, the input frame is firstly converted from RGB to YUV domain before encoding, The conversion matrix under YUV-420 format is as below

\begin{equation}\label{Eq.yuv}
	\scalebox{.95}{$
			\!\begin{bmatrix}
					Y\\U\\V
				\end{bmatrix}\!=\!
			\begin{bmatrix}\!
					0.2126&0.7152&0.0722\\-0.1146&-0.3854&0.5000\\0.5000&-0.4542&-0.0458
					\!\end{bmatrix}\! \!
			\begin{bmatrix}\!
					R\\G\\B
					\!\end{bmatrix}\!+\begin{bmatrix}\!
					16\\128\\128
					\!\end{bmatrix}\!.$}
\end{equation}
\noindent
It is observed that although luma and chroma components are obtained via different weighting factors, they are still highly correlated. Moreover, luma contains clearer textures compared with the down-sampled chroma component, so luma can potentially provide high-quality structure information to improve the chroma quality. Consequently, we extract its features to guide chroma quality enhancement.

In both spatial and frequency domain, we employ 3$\times$3 convolution layers and WBs to extract luma features and aggregate with chroma features via element-wise addition. Then, the aggregated spatial and frequency domain features are concatenated in channel-wise to produce the dual-domain Y-guided chroma feature, which is used to reconstruct the final output.

\section{Experimental Results}

\subsection{Implementation Details}

To form the training set, we compress Flickr2K \cite{timofte2017ntire} images by HM-16.7 with YUV-420 format under the all intra (AI) configuration, where the quantization parameter (QP) is 27. Then, we crop Y and U components into 64$\times$64 and  32$\times$32 small patches, respectively. Moreover, 15 videos from Common-Test Dataset for HEVC \cite{bossen2013common} are compressed under AI configuration with QP = [22,27,32,37] to generate the test set. We only train one baseline model and apply it on both U/V components and 4 QPs. Our method is implemented with PyTorch and NVIDIA GTX 2080Ti GPU. The baseline network is offline trained for 20 epochs with the batch size of 64 and learning rate of 1e-4.

\begin{table*}[t]
	\vspace{-3em}
	\renewcommand\arraystretch{1.5}
	\centering
	\fontsize{7}{7}\selectfont
	\caption{BD-rate (\%) savings for HEVC intra chrominance frames (U/V)}
	\label{table:bdrate}
	\vspace{-1em}
	{
		\begin{tabular} {c | l | cc | cc | cc | cc | cc}
			\Xhline{1.2pt}
			\hline\hline
			
			Resolution
			&Sequence
			&\multicolumn{2}{c|}{VR-CNN\cite{dai2017vrcnn}}
			&\multicolumn{2}{c|}{FE-CNN\cite{li2018fecnn}}
			&\multicolumn{2}{c|}{RR-CNN\cite{zhang2020rrcnn}}
			&\multicolumn{2}{c|}{FQE-CNN\cite{huang2021fqecnn}}
			&\multicolumn{2}{c}{OL-DN (Ours)}\\
			\hline
			
			\multirow{2}*{2560$\times$1600}
			&A1 Traffic& -3.5& -4.1& -4.2& -5.8& -11.4& -15.4& -14.1& -17.1& \textbf{-28.3}& \textbf{-38.6}\\
			&A2 PeopleOnStreet& -5.9& -5.7& -8.2& -8.6& -33.3& -29.8& -30.3& -29.5& \textbf{-46.2}& \textbf{-38.5}\\
			\hline
			
			\multirow{5}*{1920$\times$1080}
			&B1 Kimono& -1.5& -1.4& -5.0& -5.2& -20.2& -8.0& -24.0& -11.6& \textbf{-32.5}& \textbf{-38.5}\\
			&B2 ParkScene& -3.3& -2.5& -4.4& -4.1& -30.9& -6.9& -27.2& -11.1& \textbf{-40.1}& \textbf{-46.1}\\
			&B3 Cactus& -3.9& -6.3& -5.5& -10.7& -9.2& -11.7& -14.6& -22.5& \textbf{-27.4}& \textbf{-37.1}\\
			&B4 BQTerrace& -3.7& -3.0& -5.3& -6.4& -15.9& -16.4& -21.6& -33.0& \textbf{-34.2}& \textbf{-41.5}\\
			&B5 BasketballDrive& -3.3& -5.3& -10.8& -12.6& -23.3& -26.7& -25.8& -30.4& \textbf{-38.4}& \textbf{-38.3}\\
			\hline
			
			\multirow{4}*{832$\times$480}
			&C1 RaceHorses& -6.7& -11.0& -8.4& -12.5& -16.5& -23.2& -15.2& -23.4& \textbf{-21.1}& \textbf{-31.3}\\
			&C2 BQMall& -5.3& -5.3& -6.9& -7.6& \textbf{-29.3}& -32.3& -28.8& \textbf{-33.0}& -27.7& -27.7\\
			&C3 PartyScene& -4.4& -4.4& -5.4& -5.7& -17.8& -21.4& -18.9& -23.2& \textbf{-26.4}& \textbf{-24.1}\\
			&C4 BasketballDrill& -5.8& -6.8& -12.2& -14.9& -26.7& -31.8& -34.6& -41.4& \textbf{-41.4}& \textbf{-44.6}\\
			\hline
			
			\multirow{4}*{416$\times$240}
			&D1 RaceHorses& -8.5& -11.5& -9.8& -12.8& \textbf{-28.9}& \textbf{-33.7}& -26.9& -32.9& -27.0& -31.0\\
			&D2 BQSquare& -4.2& -6.4& -3.8& -6.8& -26.6& -26.5& -22.1& -29.2& \textbf{-26.9}& \textbf{-37.7}\\
			&D3 BlowingBubbles& -8.4& -7.9& -8.5& -9.0& -19.7& -23.6& -25.2& -27.7& \textbf{-25.5}& \textbf{-30.5}\\
			&D4 BasketballPass& -4.4& -6.5& -8.2& -10.3& -27.1& -29.7& \textbf{-31.9}& \textbf{-30.1}& -31.1& -28.3\\			
			\hline
			
			2560$\times$1600 &class A &-4.7& -4.9 &-6.2& -7.2 &-22.3& -22.6 &-22.2& -23.3 &\textbf{-37.3}& \textbf{-38.6}\\
			
			1920$\times$1080 &class B &-3.1& -3.7 &-6.2& -7.8 &-19.9& -13.9  &-22.6& -21.7 &\textbf{-34.7}& \textbf{-40.3}\\
			
			832$\times$480 &class C &-5.6& -6.9 &-8.2& -10.2 &-22.6& -27.2 &-24.4& -30.2 &\textbf{-29.1}& \textbf{-31.9}\\
			
			416$\times$240 &class D &-6.4& -8.1 &-7.6& -9.7 &-25.6& -28.4 &-26.5& -30.0 &\textbf{-27.7}& \textbf{-31.9}\\
			\hline
			
			\multicolumn{2}{c|}{\textbf{Overall Average}} &-5.0& -5.9 &-7.1& -8.7 &22.6& -23.0 &-23.9& -26.3 &\textbf{-32.2}& \textbf{-35.7}\\			
			\hline\hline
			\Xhline{1.2pt}
	\end{tabular}}
	\vspace{-1em}
\end{table*}

\begin{table}[h]
	\renewcommand\arraystretch{1.4}
	\centering
	\fontsize{7}{7}\selectfont
	{\caption{BD-rate(\%) results of ablation study on U/V components}\label{table:ablation}}
	\vspace{-1em}
	{
		\begin{tabular} {c c c c c}
			\Xhline{1.2pt}
			\hline\hline
			
			\multicolumn{1}{c}{Class}
			&OL-DN(w/o O)
			&OL-DN(w/o F)
			&OL-DN(w/o Y) 
			&OL-DN \\
			\hline

			A     & -21.5 / -25.0 & -34.2 / -34.7 & -11.8 / -11.9 & -37.3 / -38.6 \\
			B     & -16.9 / -16.2 & -32.0 / -34.2 & -11.0 / -16.3 & -34.7 / -40.3 \\
			C     & -20.4 / -19.3 & -27.8 / -30.0 & -13.1 / -16.3 & -29.1 / -31.9 \\
			D     & -18.7 / -17.9 & -26.6 / -30.9 & -13.0 / -15.0 & -27.7 / -31.9 \\
			\hline
			Average  & -19.4 / -19.6 & -30.2 / -32.5 & -12.2 / -14.9 & -32.2 / -35.7 \\

			\hline\hline
			\Xhline{1.2pt}
	\end{tabular}}
	\vspace{-1em}
\end{table}

\begin{table}[h]
	\renewcommand\arraystretch{1.4}
	\centering
	\fontsize{7}{7}\selectfont
	{\caption{Comparison of Increased Time $\Delta t$ at Encoder Side}\label{t.oltime}}
	\vspace{-1em}{
		\begin{tabular} {c c c c}
			\Xhline{1.2pt}
			\hline\hline
			
			Class
			&RR-CNN\cite{zhang2020rrcnn}
			&FQE-CNN\cite{huang2021fqecnn}
			&OL-DN\\
			
			\hline
			
			A &18.00\%&22.57\%&5.21\%\\
			
			B &19.00\%&23.48\% &3.60\%  \\
			
			C &18.00\%&19.77\% &2.80\%  \\
			
			D &39.00\% &19.91\% &4.90\%  \\
			
			\hline
			
			\textbf{Average} &23.50\%  &21.43\%  &4.13\% \\
			
			\hline\hline
			\Xhline{1.2pt}	\end{tabular}}
	\vspace{-1em}
\end{table}

\begin{table}[h]
	\renewcommand\arraystretch{1.5}
	\centering
	\fontsize{7}{7}\selectfont
	{\caption{Comparison of Complexity at Decoder Side}\label{t.dectime}}
	\vspace{-1em}
	{
		\begin{tabular} {c c c | c c}
			\Xhline{1.2pt}
			\hline\hline
			&\multicolumn{2}{c|}{Decoding Time (ms)}&\multicolumn{2}{c}{Increased Decoding Time Ratio $\Delta t$}\\
			\hline
			class
			&FQE-CNN\cite{huang2021fqecnn}
			&OL-DN
			&RR-CNN\cite{zhang2020rrcnn}
			&OL-DN
			\\
			\hline
			A &565&337&2947\% &24\%\\
			B &285&177&2947\% &29\%\\
			C &55&42&1667\% &22\%\\
			D &15&28&1478\% &56\%\\
			\hline
			Average&230&146&1825\% &33\%\\
			\hline\hline
			\Xhline{1.2pt}
			
	\end{tabular}}
	\vspace{-1em}
\end{table}

\subsection{Comparison of R-D performance}

Compared with other methods, the R-D performance in terms of BD-rate (\%) is given Table \ref{table:bdrate}. According Table \ref{table:bdrate}, it is found that our method outperforms the other methods for each video class, and achieves -32.2\% and -35.7\% (on average) for U and V components. Moreover, one can see that OL-DN brings more coding gains at high-resolution classes. Because high-resolution frames offer more pixel information for online training, thus improving model performance. Also, they are at larger bit rate so OL-DN extra bits occupy less portion.

\begin{figure}[h]
	\vspace{-.5em}
	\centering
	\includegraphics[width=0.35\textheight]{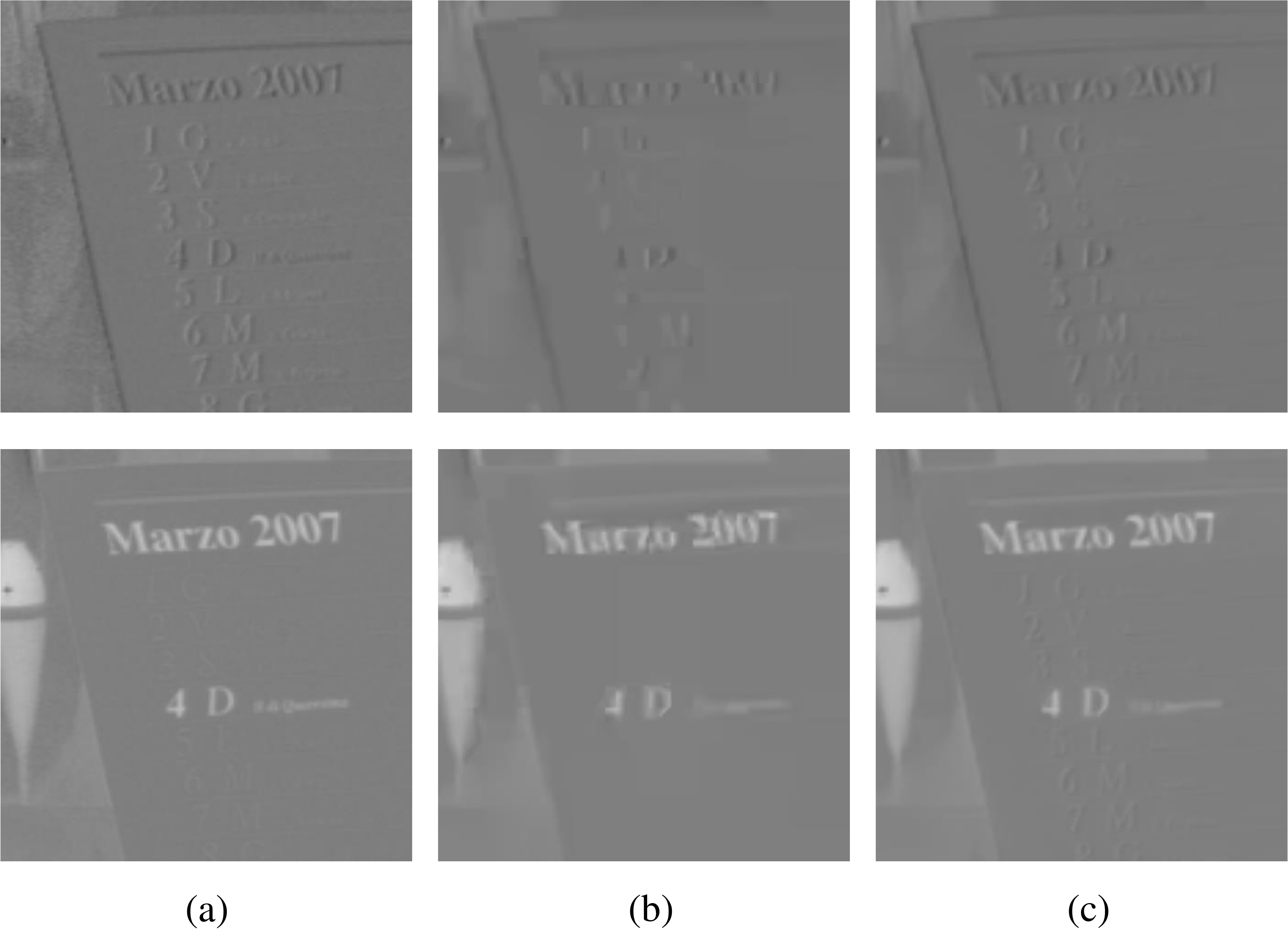}
	\vspace{-.5em}
	{\small{\caption{Visual results of \textit{Cactus\_1920x1080\_50} at QP=37. The first and second rows show Cb and Cr components, respectively. (a) raw frame. (b): compressed frame. (c): our results.}\label{f:visualresults}}}
	\vspace{-2em}
\end{figure}

\subsection{Visual Results Comparison}
Fig. \ref{f:visualresults} shows the visual results of our method enhancing HEVC compressed frames, and it is seen that our method can offer authentic textures after quality enhancement. 

\subsection{Ablation Study}
In this ablation study, we verify the effectiveness of proposed method via removing correspondent module of OL-DN, including removing online learning (w/o O), frequency domain priors (w/o F), and Y component guidance (w/o Y). The results are listed in Table \ref{table:ablation}, and they demonstrate that our methods are effective in improving quality enhancement performance.
%

\subsection{Comparison of Complexity}

Table \ref{t.oltime} and Table \ref{t.dectime} show the running time at encoder side and decoder side, respectively. The relative time is defined as $\Delta t = t'/t$, where $t'$ is neural network running time and $t$ is HEVC running time. The results indicate that OL-DN has lower time complexity compared with other methods.

%
%
%
%

\section{Conclusion}

In this paper we propose a quality enhancement network for HEVC chroma component termed as OL-DN. It online updates the model parameters to gain performance improvement. Also, we design a DCT-conv to efficiently utilize frequency priors to assist quality enhancement. Furthermore, we extract Y component feature and aggregate it with chroma feature to guide chroma component reconstruction. 

Experimental results demonstrate that this network has achieved superior performance compared with other state-of-the-art methods. Moreover, running time analysis has verified that OL-DN has acceptable complexity at both encoder and decoder sides. 

\bibliographystyle{IEEEbib}
\bibliography{ref}

\end{document}